# Attention is all they need: Cognitive science and the (techno)political economy of attention in humans and machines

Pablo González de la Torre, Marta Pérez-Verdugo & Xabier E. Barandiaran

IAS-Research Center for Life, Mind, and Society
Department of Philosophy
University of the Basque Country (UPV/EHU)

**ABSTRACT:** This paper critically analyses the "attention economy" within the framework of cognitive science and techno-political economics, as applied to both human and machine interactions. We explore how current business models, particularly in digital platform capitalism, harness user engagement by strategically shaping attentional patterns. These platforms utilize advanced AI and massive data analytics to enhance user engagement, creating a cycle of attention capture and data extraction. We review contemporary (neuro)cognitive theories of attention and platform engagement design techniques and criticize classical cognitivist and behaviourist theories for their inadequacies in addressing the potential harms of such engagement on user autonomy and wellbeing. 4E approaches to cognitive science, instead, emphasizing the embodied, extended, enactive, and ecological aspects of cognition, offer us an intrinsic normative standpoint and a more integrated understanding of how attentional patterns are actively constituted by adaptive digital environments. By examining the precarious nature of habit formation in digital contexts, we reveal the techno-economic underpinnings that threaten personal autonomy by disaggregating habits away from the individual, into an AI managed collection of behavioural patterns. Our current predicament suggests the necessity of a paradigm shift towards an ecology of attention. This shift aims to foster environments that respect and preserve human cognitive and social capacities, countering the exploitative tendencies of cognitive capitalism.

**KEYWORDS:** attention economy, engagement, habit, attention, digital platforms, enactive cognitive science, dividual

# Table of Contents

# 1. Introduction: Attention economy, platform capitalism and AI mediated extractivism

The "attention economy" is a term suggested by Herbert Simon (1971) to describe the supposedly inverse relation between information and available attention that makes the latter a scarce and valuable resource. But, having its origins in the newspapers of the 19th century (T. Wu, 2016), the attention economy now refers to a business model based on advertisement, currently in the form of advertisement agencies buying the attention captured in the most successful digital platforms (Moll, 2019; Peirano, 2019; Zuboff, 2019). The greater the captured attention, the greater the capacity for generating advertising revenue, incentivizing the technology companies that *infra*structure the attention market to design interfaces capable of attracting users' attention for as long as possible (Alter, 2017; Monge Roffarello et al., 2023; Nodder, 2013).

However, it is not enough to attract attention briefly; platforms need users to come back and stay, to get *hooked* (Eyal, 2014). At the core of this process are the strategies that developers employ to enhance the *engagement* on which platform viability depends. Engagement can be difficult to define precisely, but one can broadly say that it is "a quality of UX [*User Experience*] that is characterized by the depth of the actor's investment in the interaction; this investment may be defined temporally, emotionally, and/or cognitively" (O'Brien, 2016, p. 22). To achieve this, strategies taken mainly from the behaviourist toolkit (see section 2.2) are used by what have become known as "growth hackers", concerned with attracting (and retaining) users to digital platforms (Chen, 2012).

Besides these psychological tools, developers of digital platforms that participate in the attention economy have found an unprecedented technological ally: Big Data and Artificial Intelligence. Recent developments in this area have enabled companies to build adaptive algorithms that leverage data from user interactions to refine the designs, mechanics of interaction, and flows of the platforms themselves (Crawford, 2021; O'Neil, 2016; Roberge & Castelle, 2021) . A/B testing techniques, for instance, allow companies to test two different versions of the same feature among a great number of real users in controlled experiments to base their design decisions on their results and tune the design towards greater engagement (Gallo, 2017; Ros & Runeson, 2018). The aim is to transform the user's interaction with the platforms into a recurring pattern that keeps the participation-extraction-tuning-participation cycle in motion. Moreover, AI is playing an increasing role in corporate data-driven profiling and behavior steering capacities of digital media (Joque, 2022).

These developments take place within what has been recently labeled "computational capitalism" (Beller, 2018). The technological, social and infrastructural capacity to collect, register and manipulate astounding quantities of data left by our daily interactions is what organizes productive processes, social dynamics and what in the end produces subjectivities. Similarly, Matteo Pasquinelli suggested the notion of the "metadata society", highlighting that it is not the mere accumulation of data, but the possibility of doing meta-analysis, of mapping and interpreting patterns and trends in data, predicting tendencies and so on what constitutes a source of "cognitive capital and political power" (Pasquinelli, 2018, p. 253).

Different voices, also within academia (Beller, 2006, 2018; Carr, 2011; Citton, 2017; Crary, 2001, 2014; Crogan & Kinsley, 2012; Griziotti, 2019; Hayles, 2007; Lanier, 2018; Lovink, 2019; Rogers, 2013; Stiegler, 2012, 2019; Sunstein, 2017; Terranova, 2022; Williams, 2018), have started to denounce the perverse effects of the attention economy upon our societies and cognitive apparatuses. Some authors have even raised concerns on how the "internet of the attention economy" might be changing our cognitive capacities (mainly those related to concentration, memory, focus, etc.) overall. For Nicholas Carr (2011) the Internet has transformed its users into efficient performers of repetitive tasks requiring superficial attention, at the expense of what can be considered higher cognitive faculties. In a less deterministic vein, Katherine Hayles (2007) addressed the issue of a "generational leap in cognitive styles". She differentiated between the "deep attention" of institutionalized and stable school and pedagogical practices -associated with complex problem-solving in the framework of a single medium-, and the "hyperattention" of digital natives, a style of attention that "excels at negotiating rapidly changing environments in which multiple foci compete for attention" (Hayles, 2007, p. 188). Furthermore, Bernard Stiegler (2010) has elaborated upon Hayles' distinction, understanding the industrially mass-produced technologies of the attention economy as a form of *psychopower* that tends to displace the attentional techniques of "deep attention" towards attention to consumption.

Getting closer to psychology and cognitive science, Bhargava and Velasquez (2020) have adopted Martha Nussbaum's capabilities approach (2003) to classify the harmful effects that the excessive use of platforms is causing and construct a moral argument against this business model relying on a notion of exploitation (of cognitive vulnerabilities, reinforced by the omnipresence of the internet in our societies). Castro and Pham (2020) introduce into the equation the concept of 'noxious markets', drawn from Debra Satz (2010), enabling them to build an argument in favour of platform regulation, similar to what was done with tobacco in the past. What both approaches share is the recognition of the harmful and addictive nature of attention economy platforms. In fact, it is starting to be shown that patients with problematic patterns of internet use show damage to the brain's dopaminergic system, similar to what is found in other kinds of addiction (Hou et al., 2012).

For sure, this is not the first time that attention has become a battlefield. The process of modernization -with its intertwinement of social changes, scientific theories of the organisation of labour, the birth of a proto-mass culture and the emergence of experimental physiology of perception in the second half of the XIX century-, established attention as a political issue through wich "the perceiver becomes open to control and annexation by external agencies" (Crary, 2001, p. 5). We are witnessing now a new alliance between a psychological theory (behaviourism), enhanced with the possibilities that Big Data and machine learning allows, and the corporate giants that control platform capitalism (Srnicek, 2016). This has also brought a new reorganisation of labour, social patterns of interaction and power in the era of extractivist cognitive capitalism (Pasquinelli & Joler, 2021). As the affective, social and communicative capacities of human beings became integrated in the production of economic value (Barandiaran et al., 2024; Moulier Boutang, 2011; Vercellone, 2006), a new kind of psychopolitics (Han, 2017) emerged here, marking the transition from the so-called disciplinary societies and their biopolitics (Foucault, 1991) to contemporary societies of control (Deleuze, 1992). We are no longer in the era of the government of individuals or populations, but in the time of the control and modulation of fragmented subjects, *dividuums* (Raunig, 2016) and their networks emerging from the extraction and treatment of their data by the powers of the platforms that articulate the attention economy.

We still need, however, to find a scientific approach that can accommodate an accurate description of what it means that our attention is captured by digital platforms, of what it means to govern through the fragmentation of the individual, while also providing a normative standpoint from which to understand the harmful character that this attention capture can have from a naturalized perspective. Mainstream approaches to the scientific characterization of attention, theories of what attention *is*, provide allegedly "neutral" or non-normative accounts of its contribution to personal autonomy. The negative effects of capturing attention appear thus as something contingent or additional to the scientific understanding of attention. On the contrary, we shall argue, it is possible and necessary to characterize attention in a manner that makes its capture inherently detrimental, evidencing the privatizing-depriving dialectic involved in the workings of platform capitalism.

Here is where the fruitfulness of our approach, that will be based on the enactive notion of habit, comes to force. 4e-cognition approaches emerged in the last decades as a reaction to the limitations of the cartesian intellectualist and rationalistic presuppositions (Newen et al., 2018; Wheeler, 2007) that pervaded mainstream cognitive science. Beyond representation, computation and information processing -building blocks of the cognitivist revolution (see Fodor, 1980; Newell & Simon, 1976; Putnam, 1965)-, this family of approaches emphasized the role in cognition of the body (Gallagher, 2023; Shapiro, 2019) the situation (Hutchins, 1995), the equipment (Clark, 2003; Clark & Chalmers, 1998) or the organisation of the dynamic sensorimotor interactions that constitute the mind (Di Paolo et al., 2017; Noë, 2004; Thompson, 2007; Varela et al., 1991). Mind and cognition thus leave the realm of the isolated brain, and the environment becomes a constitutive and central element in their study. Beyond the "human-as-information-processor" framework, cognitive processes such as attention can be studied under a richer light, analyzing how "cognitive extensions" (Choudhoury & Slaby, 2016) organize and shape them. This is what allows to capture their strong technopolitical dimension, naturalizing in a scientific study of attention the potential harmful effects of certain environments.

We will devote section 2 to explaining why classical approaches from cognitive and behavioural sciences lack an adequate analysis of the phenomenon. We will start in section 2.1 by reviewing psychological theories of attention and later on expand our review in section 2.2 to behaviourist techniques used to craft *engagement* in digital platforms. Based on this, we will conclude that we need accounts of attention that understand it as an emergent psychological process of an agent that behaves within a particular environment, but that we are still missing a naturalized normative standpoint from which to judge the effects of designed engagement. This will lead us to present in section 3 an alternative approach based on the enactive conception of *habit*. Indeed, it is the notion of habit that is at work in the change of cognitive styles denounced by Carr, historicised by Hayles and politicized by Stiegler. An account of users as agents understood as webs of networks of habits grants us a normative standpoint from which to judge the harmful character of certain habits. This will allow us to explain how successful cases of attention or engagement can nevertheless be harmful cases of attention or engagement, also to derive some conclusions for political philosophy and to underscore the relevance of a political philosophy of natural and artificial intelligence.

## 2. Attention & engagement

### 2.1. From the cognitivist bottleneck towards coherentist theories of attention

Attention became a central concept in the middle-of-the-century turn away from behaviourism and towards cognitive psychology: the focus on the cognitive subject as an information-processor with limited computational capacity made attention into a central and necessary selective process. The main proponent of this view was Donald Broadbent, for whom "a nervous system acts to some extent as a single communication channel, so that it is meaningful to regard it as having a limited capacity" (Broadbent, 1958, p. 297). The idea is that these limitations in the management of perceptual information are concentrated on a critical bottleneck at the intersection of two processing systems operating in series (the first is able to absorb a large amount of stimuli and operates automatically, while the second, more limited in its capabilities, needs to be deployed selectively). The bottleneck resulting from the connection of these two subsystems corresponds to attention, in the sense that "when a representation of a stimulus passes through that bottleneck, the stimulus ipso facto counts as one to which attention has been paid" (Mole, 2021).

This is the context where the classical idea of the attention economy as proposed by Simon arose. The broad idea is that an abundance of information in the environment contrasts with a relative scarcity of available attention from the nervous system. For instance, Atchley and Lane (2014) attempt to tackle the effects of the attention economy from a cognitive science perspective, and use for this an approach based on processing limitations and the idea of incurring in an "attentional debt" when many things fall outside what can get through the filter of attention. If that is the case, many of the tasks that need higher order cognitive processing will have to be managed only with the superficial and automatic processing resources of early-stage processing, thus worsening performance. Our brains, according to this view, adapted to simple environments, are now being affected by multiple stimuli from a tremendously information-rich environment in the context of digital platforms. The digital contexts of the attention economy may be worsening cognitive performance "simply by placing more processes in the queue" (2014, p. 138).

However, this cognitive approach rests on some difficult assumptions: for instance, what exactly does it mean that the amount of information in the environment has increased with digital devices, if we are to understand the cognitive agent as a mere abstract information processor? The digital environment has no more information for a subject who does not know how to surf the Internet (or what a computer is), than the jungle does for someone who doesn't know how to navigate it. We therefore understand that Atchley and Lane must be referring to an increase in the *relevant* information for the agent. But what determines this relevance remains underspecified, making it depend on the normative assumptions that the authors introduce into the equation without justifying them. We need to specify the relevance of information in the environment in relation to the organization of the agent that inhabits it. Atchley and Lane, and Broadbent and Simon before them, offer an internalist, brain centered view of agents as passive information processors overwhelmed by the characteristics of an external and independent environment. Their theories seem incapable of capturing the interplay of mutual influences between environment and agent. Also, the "limited resource" metaphor, has been argued, seems to be merely a redefinition of the explanandum (Navon, 1984; Neumann, 1987a): postulating selective mechanisms to explain how an information bottleneck is overcome is tantamount to saying that selective mechanisms create a bottleneck. Furthermore, that metaphor would imply the existence of an upper limit of processing, whose search has been unfruitful in the laboratory. What has been found instead is "simply a trade-off, or interference (of whatever kind) in the ability to perform temporally overlapping tasks" (Allport, 2011, p. 29).

All of this led to a point where, if the study of attention in its early days was dominated by computational notions of information and processing capacity, from the 1980s onwards bodily and interactive characteristics of agents entered the picture, leading to a notion of attention more linked to agency that crystallized in the so-called selection-for-action theory (Allport, 1987). For Allport, it is the broader "dynamic process of unification--both integration and segregation of neural activity--that is the relevant causal process, and attentional phenomena, including their limitations (bottlenecks, limited processing capacities) are their observable behavioural consequences" (Allport, 2011, p. 32). Ulric Neisser (1976) had paved the way for these approaches by questioning the centrality of the brain in information processing limitations. Through a series of experiments, he was able to show how the supposed bottleneck had less to do with processing limits than with behavioural coordination requirements. It is bodily limitations that prevent us from looking in two directions at the same time, or playing the guitar and lighting a cigarette simultaneously. The selectivity of attention was therefore about the requirements that our corporeality imposes on our cognitive processes. Allport (1987) and Neumann (1987b) generalized this coherentist principle to the totality of cognitive processes. It is the need to maintain the coherent course of an end-oriented action that explains the selective role of attention in our mental life. Later on, Wayne Wu built on these results to develop a philosophical theory of attention based on the ubiquity of what he calls "the many-many problem" (W. Wu, 2014). The need to select among multiple sources of information, and to select among the possible things that can be accomplished with that information, is for Wu, the problem that attention is charged with solving.

For Desimone and Duncan, working from behavioural neuroscience, it is the existence of competition between stimuli that struggle to gain control of the agent's activity in the context of task performance that calls for selection. Their model, called the "biased competition hypothesis" (BCH), explains the different attentional processes, at the personal and subpersonal level, as a result of the numerous struggles between different stimuli. For the specific case of visual attention, attention appears "as an emergent property of many neural mechanisms working to resolve competition for visual processing and control of behaviour" (Desimone & Duncan, 1995, p. 194). In Duncan's words, "coherent attention develops as different systems converge to work on related cognitive content" (Duncan, 2006, p. 2). Salient stimuli can ""win" this competition and dominate the system in a "bottom-up" fashion, simply because the neural response elicited by them is much stronger than that associated with any other object" (Ruff, 2011, p. 5). This is what neuroscientists call stimulus-guided or exogenous attention, which is concerned with explaining why perceptually salient stimuli, such as a loud noise or garish colours (or a sudden bright and white pop-up box on top of your screen when a notification arrives), are difficult to ignore. Each of these competitions is, in turn, biased by "top-down" processes linked to the task that the agent is facing, her goals and expectations, in what is known as "endogenous" or "voluntary" attention. I might not see any bright notification lights in the upper corner of my phone, but still check my phone screen on the table just in case. Besides these two "modes" of attention, the BCH model describes a winner-takes-all principle; given the interconnectedness of different neuronal populations, "enhancement of activity devoted to a given object---either bottom-up (...) or top-down (...)---in one area should also influence activity in all corresponding interconnected neuronal populations" (Ruff, 2011, pp. 5–6). In sum, what Desimone and Duncan offered was a set of identifiable neurological mechanisms from which attention emerges as a global psychological process. There is no central agency responsible for attentional processes in this model, but only the very processes of competition between stimuli, some of which are not even attention-specific processes. Coherence is necessary across scales, to avoid the risk of neural, perceptual or behavioural chaos.

Desimone and Duncan's framework is still widely used to think about attention. However, Christopher Mole (2011) has proposed a philosophical reading of the BCH model in which attention is not identified with the neurological processes that underlie it, but would have more to do with the nature of "unison in an orchestra". He considers that the BCH correctly tells us how "many instances of this attention-realizing unison come about, and of how they get maintained" (Mole, 2011, p. 133), but not what attention *is*. Attention requires the synchronized functioning of the processes from which it arises, but is not identified with them, as these may vary according to the task at hand. In Mole's reading, for an activity to be attentively performed, no resources relevant to the activity must be engaged in unrelated tasks. This reading of attention emphasizes the global and personal character of attention. This view is close to enactive readings of attention as an "agent-level phenomenon of the task-relevant, cognitive processes operating in unison" (Thompson, 2020, p. 78). Furthermore, the notion of affordance from the tradition of ecological psychology, as "the possibilities for action that are available to agents in their environments" (Heras-Escribano, 2019, p. 3) can also become useful when understanding attention from a broader agent-level perspective. Both the structure of the environments and the organisation of the agent are relevant for characterising the system of affordances. Through "their insertion into a particular socio-cultural context, human beings develop concrete bodily skills and abilities" through what Gibson called the "education of attention" (Gibson, 1979, p. 254).

The coherentist position, in sum, reverses the sign of the economy of attention as it has been traditionally understood. In contrast to the management of limited resources posited by bottleneck theories, what we now need to manage is excess capacity. Because human agents can process multiple stimuli, they need a mechanism responsible for avoiding the capture of their attention by distractions, while maintaining the coherence of their actions. From this point of view, attention provides coherence and avoids interference between the activities of an agent with the capacity to process more stimuli than those relevant to its actions. This calls for more nuanced analyses of how attention works, than that outlined by Atchley and Lane's from a purely cognitivist perspective. We need to account for an agent and her behaviour, the task she is facing and the environmental conditions upon which she is doing so, to understand how attention brings about a specific course of action. Furthermore, Atchley and Lane's perspective overlooks the deliberate nature of the design of the attentional capture mechanisms posited by digital platforms. In fact, developers do design their platforms with a broader behavioural perspective in mind, as we will see in the following subsection.

## 2.2. Designing for engagement: capturing attention

One of the key concepts in the attention economy is that of *engagement*. Engagement has a global and behavioural character; what developers are interested in is the users spending as much time as possible interacting with the platform, not merely in certain stimuli passing through a supposedly attentional bottleneck. That is, on users generating a *habit* of interacting with the platform. How to achieve this has been an object of interest and research for quite some time now. As one milestone of this research, B.J. Fogg presented a model for persuasive design where he stated that "for a person to perform a target behaviour, he or she must (1) be sufficiently motivated, (2) have the ability to perform the behaviour, and (3) be triggered to perform the behaviour" (Fogg, 2009, p. 1). Working upon this, more recently Nir Eyal (2014) presented a four-step model for building "habit-forming products", where the key focus points are: triggers, actions, rewards and investment. The takeaway from these models is easy: in order to get their users to perform an action one should (1) establish a trigger that is gradually internalized by them, (2) make the desired action as easy as possible to perform, (3) reward its performance (intermittently for maximum effect), and (4) work on creating (mostly social) motivation for the users to continue performing the action. Let us delve a bit further on each step.

Under these approaches, a big part is setting up *triggers* that prompt certain behaviours. This kind of triggers can be understood as *discriminative stimuli* that get associated with a behaviour that gets rewarded, thus making the emitting of that behaviour more probable[1]. The classic kind of triggers are salient, hard-to-miss, stimuli in the environment. These are the stimuli that, if we use BCH terms, are designed to "win" the bottom-up competition for neural processing. The light we see in the upper corner of many smartphones, indicating a notification, is an obvious example of a trigger: it signals the behaviour of checking our phones, which will be rewarded. This is the first condition to set up a behavioural habit. However, any design that depends on users being constantly in perceptual contact with the device (so that they can see/hear the designed trigger) can only go that far; the real challenge is to create habits that prompt users to use a device when they are not even in contact with it. Eyal (2014) strengthens in his book that the best route for success in building habit-forming products is to make people see the product as a relief for feelings of discomfort. This is also the case for behavioural addictions in general, as Alter (2017) notes. The question here is managing to get the very feeling of discomfort (either stress, anxiety, or plain boredom) to function as a trigger for device or app use. *Internal* triggers, in combination with good and salient external triggers, are the key to securing the frequent initiation of a desired behaviour.

---

[1] This is different from the reflex-like association of stimulus-response in classical conditioning, where behaviour is not emitted but *provoked* directly by a conditioned stimulus. Under operant conditioning, triggers (discriminative stimuli) precede a behaviour that is then rewarded, so that a three-factor association takes place (Skinner, 1965).

Another ingredient in making a behaviour occur as probably as possible, is making said behaviour easy to enact. The simplification of interfaces during the last decade and the overall ease of use of most current platforms is not coincidental, but a way in which designers reduce the skill needed to perform an action -or the amount of micro-actions required to do it. This can be appreciated in the reduction in configuration and customization possibilities of platforms, where many actions are being partly and conveniently "done for us" without our needing to even think about them. For example, we are used to the fact that icons of the applications we download appear by default on our smartphone's desktop for easy access. Similarly, the next episode of the series we are watching on streaming platforms plays automatically without the need for us to press any buttons. And as one logs in on certain social video platforms, content appears in the home page, without us needing to search for it, requiring us only to move our thumb up and down to scroll through it. Platforms are designed in such a way that we encounter no difficulties when using them, favouring a quick acquisition and stabilization of the behavioural response associated with the triggers we have established (for further discussion, see Pérez-Verdugo & Barandiaran, 2023).

Once designers have set a trigger, and facilitated the performance of a certain behaviour, they need to reward it. Behaviourism has carefully explored for decades how differences in contingencies of the environment act as rewards or punishment for human beings. Whenever something acts as a reward for someone, he or she is more likely to repeat the behaviour that is associated (most basically because of time/space contiguity) to that reward; that behaviour is reinforced. However, it is also known that if we merely reward every instance of a behaviour, said behaviour will die out as soon as we stop reinforcing it, as it will undergo "extinction" (Skinner, 1974, p. 64). The most effective way to get someone to continue performing a certain behaviour repeatedly and steadily in time, even when we have stopped reinforcing it, is through a schedule of variable rewards, what is known as intermittent reinforcement (Skinner, 1965). This means that sometimes the behaviour gets rewarded, and sometimes it doesn't, in an arbitrary fashion. Intermittent reinforcement modulates the release of dopamine, a neurotransmitter generally associated with pleasure and rewards, but also to pre-reward motivation (that is, to anticipation) (Wise, 2004). Digital platforms exploit this intermittence continuously: whenever we are scrolling down on short videos, we never know whether the next one will make us laugh or whether it will be a dull video of no interest for us. Similarly, when we post a picture of ourselves, we never know whether we will get a decent amount of likes or whether the picture will "flop" this time (this has been particularly enhanced with recommendation-based feeds, where this arbitrariness is mediated by algorithms that make your posts more or less visible for others).

Intermittent reinforcement is a very powerful behavioural tool to motivate people. However, digital designers use rewards (and punishments) in more intricate and complicated ways to get people to perform their target behaviours. Eyal (2014) highlights in his model that one of the main motivators is our past investment in the platform: the more time and resources we have already invested in a platform, the more motivated we will be to continue using it. As Nodder (Nodder, 2013, Chapter 1) also states, this can be seen as a way to resolve the "cognitive dissonance" created by our having spent time in the platform without necessarily wanting to. Besides investment, the social dimension of many of the platforms is a goldmine for motivating people (not coincidentally, many platforms that wouldn't necessarily be social in nature include a social feature for motivating users, some with arguably good intentions such as hiking or language learning apps). Social conformity and social proof phenomena motivates us to behave as we see others behave, and app designers take good care in showing us what other people like and do, for example, in social media apps.

As these models point out, the engagement behaviours that stem in the attention economy are much more complicated than what focusing on attention as a mere bottleneck in an information-rich environment would grant us. We are not merely abstract processors that can only attend to a small part of the many stimuli of our environments, nor can we focus only on specific neural responses that grant the coherence of our actions; our behaviour as a whole is being modelled to get this attention. Getting past the focus on attention as information-selection calls for a broader perspective that takes behaviour as a whole, without assuming a binary split between perception and action.

Behaviourist ideas, if complemented with neuroscience of attention (and perception, and action) in a systemic way, can offer such a broad picture through the notion of *habits*. However, we are still missing a fundamental aspect: the ethical and political question of *what it means for the agent* to engage in those habits. Attention grants coherence to our behaviour.

But, to what behaviour? It may well be that the coherence of our actions is being harnessed to generate huge revenues for the big corporations of cognitive capitalism while our psychic and social processes enter a permanent crisis. It is the ability of platforms to capture attention that solves Wu's many-many problem in a taxing way: from among the multiple possibilities for action, hoarding or consumerist activity is encouraged. Engagement behaviours are successful habits; the cognitive unison that is attention works perfectly in giving coherence to them. But as we reviewed in our introduction, while undoubtedly successful cases of attention and engagement for platform developers, these habits are not always successful habits *for the agent*. The approaches just reviewed don't allow us to make this jump in a justified way. The absence of a normative dimension linking identity and behaviours of the subject does not allow us to naturalize the idea of a perverse habit of engagement. We need to zoom out a bit more and provide a more encompassing account of the agent that performs said habits. Enactivism, and its conception of agents as webs of networks of habits, grants us such an account.

## 3. Sensorimotor autonomy and the attention economy

### 3.1. From the behaviourist to the enactive notion of habit

Behaviourism belongs to a long tradition of thought that has seen habits as an organizing principle of the human psyche. After the decline of behaviourism and the simultaneous emergence of cognitive sciences, the concept virtually disappeared from psychological research, replaced by the concepts of representation and computation (as we could see in early cognitivist views of attention). Behaviourism, as part of a broader associationist tradition, conceives habits atomistically in terms of statistically significant relationships between stimuli and responses (triggers, behaviours and rewards), disregarding the mental processes involved in habitual behaviour and presenting habits from an automated and rigid conception of human action (Barandiaran & Di Paolo, 2014).

There is another tradition, however, for which habits are a transversal phenomenon capable of encompassing different strata of an agent's life. With its remote origins in Aristotle, and modern and contemporary figures like Hegel (2017), Ravaisson (2008) , Dewey (1957) or Merleau-Ponty (2006), so-called organicism tends to embrace a more ecological and self-organizing perspective, relating habits "to a plastic equilibrium that involves the totality of the organism, including other habits, the body and the habitat they co-determine" (Barandiaran & Di Paolo, 2014, p. 5). In contrast to associationist atomism, organicism maintains a holistic version of habits as networks of self-organized processes that maintain relationships with each other in need of adjustment and composition. Enactivism (Di Paolo et al., 2017; Thompson, 2007; Varela et al., 1991), heir to this tradition, conceives habits as a self-sustaining form of life characterized by its adaptive and precarious nature, with its own normative conditions of persistence (Barandiaran, 2017).

Nested within the autonomy of the living organization directed at biophysical self-maintenance and reproduction (Ruiz-Mirazo & Moreno, 2004), new forms of normativity emerge at the level of sensorimotor dynamics. This normativity derives from the habitual nature of the actions, which generate self-sustained sensorimotor patterns with characteristics analogous to biological systems in that their identity is linked to their own conditions of execution (Barandiaran, 2008; Egbert & Barandiaran, 2014). Habits then, defined in these sensorimotor terms, depend on different "support structures" situated along the brain, body and also the environment (Di Paolo et al., 2017). Thus, smoking may be altered after the birth of a daughter, which often limits the possibilities of smoking in the domestic context. This change in the environment may, over time, disorganize the network of processes that shape the habit, ending the smoking behaviour of the new parent. The neurodynamic patterns that underlie habitual behaviours are therefore sustained by sensorimotor activity, which in turn depends on a certain stability of the environment that sustains (and is shaped by) habitual actions. Habits are operationally closed adaptive and precarious structures of neurological, bodily and interactive processes that give rise to behavioural patterns with their own normative requirements (Egbert & Barandiaran, 2014).

The notion of habit offers us the flexibility and plasticity necessary to understand mental processes from a normativity that cannot be reduced to biological requirements. Following the intuition of William James (James, 1995), halfway between the organicist and associationist traditions, organisms can be understood as networks of habits. Thus, in addition to the normativity of each habit, networks of habits also have conditions of existence that require the maintenance of the coherence that sustains the agent's identity, and both normativities can enter into relations of confrontation (Barandiaran, 2008). This new normative standpoint, different from that of the single habit, that emerges when we understand agents naturalistically as networks of habits, is what provides the adequate grounds to analyse how successful cases of habit formation (or attention capture) can nevertheless go against the viability of the agent (see also Pérez-Verdugo & Barandiaran, 2023). The behaviourist notion of habit that organizes the economy of attention can therefore be understood as a particular and decontextualized case of the network of habits that configure the agent's identity, one more habit that enters into complex relations with the rest of the habits. When one habit endangers the integrity of the whole network (the identity of the agent), we are in the presence of a "bad habit" (Ramírez-Vizcaya & Froese, 2019). When the habitual attentional patterns of the attention economy endanger the recurrence of the network of habits in which the agent identity consists, we find ourselves in the presence of a sociotechnical produced bad habit[2].

## 3.2. The habit of engagement and its sociotechnical production

With the notion of habit enriched by enactivism, we can try to naturalize attentional capture as a network of complex processes that give rise to engagement, understood as an emergent process that involves neurological, bodily and environmental aspects. In this picture we can identify (as we sketched in section 2) neurodynamic, interactive, bodily and social patterns that play a fundamental role, together with that of the digital environment specifically crafted with the aim of inducing and stabilizing certain habits of the agent.

We are interested in a notion of attention, as the one offered by Allport, as an emergent psychological property understood as a selective relationship between an agent and its environment, realised at a agent-systemic level "by a corresponding relationship among brain, body, and environment, including an appropriate, integrated *whole-organism brain state*" (Allport, 2011, p. 26). We believe that this emergentist definition allows us to connect with the enactive naturalisation of the processes of attention capture that occur in the attention economy. It takes up the idea of the selective nature of attention, underlines the coordinating role of attention in our behaviours, and also allows us to study the influence of causal processes flowing in various directions in the origin of mental phenomena (Thompson & Varela, 2001). This definition in turn allows us to integrate all other subpersonal meanings, such as processing constraints or competing neural processes, as manifestations of the primary experiential level of attentional phenomena. A situated, embodied, interactive and distributed reading of attention assumed from the enactivist theory of mind allows us to introduce the necessary normativity to evaluate the attentional patterns to which the designs and functioning of the digital infrastructure of computational capitalism may give rise. Attention encompasses different scales, ranging from competition between the purposes of different neighbouring neurological subsystems to the characteristics of the environment, social and cultural norms or evolutionary demands.

Attention as a global phenomenon is explained by large-scale neuroscience as a pattern of interactive activation between different brain areas through synchronization processes (Allport, 2011) allowing for a coherent course of action. The stabilization of a certain pattern of neural interaction is mediated by the dopaminergic system that is being modulated through the intermittent reinforcement that users encounter in platforms. This functions as a Hebbian learning process that fixes habitual patterns of activation involved in the use of those specific digital platforms. Furthermore, with the stabilization of said patterns, a certain base for top-down attentional processes focused on the specific habit (task) we are dealing with, starts to form. As we saw, this is one of the goals of designers: to get from external triggers reclaiming neural responses in a bottom-up fashion, to internal triggers that bias these responses top-down. Once this is established, it is also easier to keep playing with design. For instance, it has been shown that when focused on a task, distractor (non-related) stimuli that fall outside our current focus of attention are more difficult to ignore when they are perceptually similar to the stimuli relevant for the task, as compared to other (equally salient) distractors that share no similarities with task-relevant stimuli (Ruff, 2011, pp. 5–6). In sum, the current task we are dealing with biases what kind of salient stimuli "grab" our attention, and this creates a reinforcing cycle that strengthens the related neurodynamic habitual patterns. In digital design, this is particularly notable in a certain brand of "dark patterns", related to effects of "interface interference" -where an interface is manipulated to favor certain actions over others- (Gray et al., 2018), particularly in disguised ads that are made to look similar to the content you are supposedly interested in interacting with (Monge Roffarello et al., 2023). If I am trying to buy a plane ticket, any bright green slightly-round shaped button with an all-caps word inside it will be perceptually hard to miss. This effect cannot only be explained in terms of a bottom-up perceptual salience; it is the relevance for the task that the agent is trying to perform that explains why it is so attention-grabbing, and why it might not be for another person who is not buying tickets online.

---

[2] Ramírez-Vizcaya & Froese's (2019) account centers around addiction; we have already highlighted in the introduction how many analyses of the dangers of engagement bring it close to phenomena of addiction.

The more we enact a network of habits, the more robust said network will become, and the harder it will be to change it. The more I enact the habit of scrolling on TikTok on my phone when faced with a blank document in my computer, the stronger the habit of scrolling on TikTok to avoid work will become. As noted by Jesper Aagaard, these kinds of phenomena should be seen as *habitual*, because they are "explainable neither in terms of mental choices nor mechanical reactions to stimuli, but as deeply sedimented relational strategies" (Aagaard, 2015, p. 95). In general, our behaviour within digital platforms cannot be understood without taking into consideration the technological environment, the bodily agent situated in it, and their mutual relations across time. The enactive habit as a unit of analysis allows us to take all of this in consideration when analyzing patterns of engagement.

Platform engagement habits can thus be conceived as *a network of neurological, behavioural and techno-ecological processes that sustain patterns of behaviour recurring over time and endowed with their own normativity*. The balance between this series of processes is what constitutes the robustness of the habits of the attentional economy. The attentional capture processes designed by platform developers induce and stabilize the network of interactive and neural processes that underlie the habit of engagement; on which the advertising-based business model of the attention economy depends.

The engagement process is a recursive one, where the use of the platform generates the data that will be mined and analysed to generate more engaging interfaces that maximise the time users spend on them. As Alter points out, "the people who create and tune technology (…) do thousands of tests with millions of users to understand which tweaks work and which don't, which background colours, which fonts and audio tones maximise engagement and minimise frustration. As the experience evolves, it becomes an irresistible weaponised version of its former self. In 2004, Facebook was fun; in 2016, it's addictive" (Alter, 2017, p. 4).

## 4. Discussion and Conclusions

### 4.1. Artificial Scarcity: an integrated account of attentional capture through habits

As we have seen, cognitivist assumptions favour an intellectualist, brainbounded internalist conception of attention that conceal both the sensorimotor and environmental aspects of the attention phenomena and the deep precariousness of human agency (that grounds its normativity). As rational agents that can process limited information, so we are told, the market competes for optimizing attention, to bring in specific information to our scarce processing resources. Enactivism makes it possible to invert the equation. The scarcity is not the cause, but the lived effect of the attention economy. If attention is undestood as an extended phenomenon, involving the coordination of brain-body-environment components, then the marketing of attention (i.e. the integration of our cognitive capacities in the market production of economic value) can be thought of as the deliberate process of externalizing it as a commodity. That means *privatizing* attention, *depriving* the subject from attention in the process of making it an external commodity. It should also be noted that the very conception of attention as a scarce resource, given its economicist character, is not innocent when trying to analyze the attention economy of digital contexts. As Terranova argues, it is a way to reintroduce economicist ways of thinking in a medium, the Internet, in which until now we had spoken of abundance . Through the scarcity model, "the wealth of information creates poverty that in its turn produces the conditions for a new market to emerge (…) (that) requires specific techniques of evaluation and units of measurement" (Terranova, 2022, p. 29).

It is not, then, that the attention economy has a set of secondary side effects on the subject: the privatizing-depriving process *is* the primary effect. An embodied and enactive conception of habit, its precariousness and its constitution through environmental "support" structures, makes it possible to re-conceptualize attention from a techno-economic point of view. Cognitivist theories of attention, by focusing on computational and/or neural intra-cranial mechanisms of attention, conceal the fundamental role that environmental factors can and increasingly play on structuring both attention and behaviour.

Our cognitive abilities depend not only on our body and brain, but develop in a natural, social, technological and cultural environment, subject to certain norms and expectations. This environment of cognitive institutions has a constitutive role on our mental life. The socio-cultural environment of institutions "plays an essential role in the formation of an embodied subject's characteristic patterns of attention, engagement and response" (Maiese & Hanna, 2019, p. 54). The institutional contexts in which this capacity to regulate behaviour unfolds have a direct influence on its exercise. The attention economy not only surveils but crucially "surwills" (Barandiaran et al., 2024; Calleja-López et al., 2018) its users, that is, shapes and organizes, facilitating the generation of harmful habits through attentional capture, jeopardising the coherence of the subject's sensorimotor autonomy. The attention economy is turned into an intention economy.

The political dimension becomes explicit from this theoretical perspective. Being aware of the role of novel technoscientific developments in the constitution of our mental lives, in the form of the environmental support structures that scaffold our habitual behaviours, allows us to recognise who are the "designers" of our experience and with what effects.

## 4.2. Dividual management and the dismantling of habit integration

And this brings us to the necessity of putting our attention on the dynamics of this concealed environmental dimension, the social and political side of our mental constitution. Digital platforms powered with AI techniques should be understood in the economic and civilizational paradigm of extractive cognitive capitalism, in which knowledge, information, affects and social relations become essential for the production of value (Moulier Boutang, 2011; Pasquinelli, 2023; Vercellone, 2006). This *apparatus* (McQuillan, 2022, p. 146), "a layered and interdependent arrangement of technology, institutions and ideology", combines specific subjectivation effects with specific technopolitical affordances. If we follow Deleuze´s periodization of societies in terms of the machines that better characterize historical periods, we are moving from foucauldian *disciplinary societies* (illustrated by thermodynamically driven mechanical industry) to informationally driven *societies of control* (Deleuze, 1992). Our approach adds to the equation a new understanding of how the modern notion of individual gives way in the societies of control to the *dividual* (Deleuze, 1992; Han, 2017; Raunig, 2016), a product of the technical possibilities that cognitive capitalism exploits. When statistical profiles are generated (by data extracted from massive user behaviours) and used to re-desing interface environments to induce and stabilize specific behavioural patterns, the subject appears disaggregated, divided into externally managed behavioural chunks. It is not the human person, but AI system, by creating profiles and inducing behavioural schemes, that fundamentally organizes behaviour in digital platforms.

The passage from governance of the individual to the modulation of *dividuals* is made explicit through the sociotechnical production and management of habits on the platform economy. This new form of psychopolitics (Han, 2017) calls for new ways of understanding the constitution of the mind, one able to underscore the sensorimotor and precarious nature of our mental life, able to incorporate the multitude of processes and influences occurring between brain, body and environment. Crucially, the enactive notion of habit, with its externalist virtualities, allow us to naturalize the notion of the *dividual:* it can be understood as each of the habit-subnetworks that constitute the identity of the subject and that are supported (and reinforced) by a dynamical hyperdesigned adaptive environment through deliberate sociotechnical processes with political, economic and psychological effects. If oppression has been thought of as associated with technologies of subjection, we now find the opposite movement of active repression: the dismantling of the individual, in a process of technological pulling-apart of the integrative capacity of the complex web of habits constituting a person. This is perhaps why non fiction top bookshelling lists are full of habit centered self-help manual (Clear, 2019; Duhigg, 2012; Tynan, 2014; Wright, 2020), in a desperate attempt to regain ownership of (healthy) behavioural patterns.

### 4.3. *AI* attention is all they *will* need?

The era of human attention-deficit is the era of machine attention-abundance. This coincidence requires, itself, some attention. The new-AI revolution that has given birth to ChatGPT, and a myriad of Large Language Models, comes fuelled by two main factors: 1. The huge investment on computational resources: processing capacity and training data, and 2. A new architecture for computational attention: the transformers, with the now widely celebrated paper "Attention is all you need" (Vaswani et al., 2017) that has boosted the success of Large Language Models and multimodal deep learning generative architectures. Interestingly, the new capabilities opened by these systems have fuelled the proliferation of *assistants* and so-called "autonomous agents" (Andreas, 2022; for a critical evaluation see Barandiaran & Almendros, Forthcoming; Wang et al., 2024; Xi et al., 2023).

An increasing reliance on artificial assistants on the performance of digital tasks might likely displace the attention capture to artificial systems (as well as humans). The algorithmic violence for attention is not new. It suffices to look at how cognitive capitalism has already seriously eroded attention-proxies like Google Search. Search engines are a major piece of attention organizer, the filter an orgazine potential digital interactions and direct behaviour to certain sites. It is an increasing reality that search results have seriously degraded (Bevendorff et al., 2024) due to massive marketing investment to outrun competitors on search engine "attention" capture (instead of content quality) as a proxy for human attention. But AI-populated digital environments might drive this violence to a new level. As digital assistants become the next step in the externalization of attention, so will the battle for their capture, and their recruitment for that task. The hybrid attentional battlefield is now difficult to imagine. Not only will the sensorimotor digital landscape be increasingly patched by dark-patterns, but also increasingly populated by dark-agents. In order to preserve some behavioural autonomy, we might well need to navigate the web with attention guards that actively counteract artificial attention-capture agent, (as we already, partially, do with attention shields blocking pop-ups and ads on our browsers).

## 4.4. The need for a political philosophy of (natural and artificial) intelligence

The approach we have developed allows us to exercise critique from a perspective that combines tools from cognitive science, philosophy of mind and the critical analyses of political philosophy. As the emerging field of political philosophy of mind (Gallagher, 2013; Maiese & Hanna, 2019; Protevi, 2009, 2022; Slaby, 2016) shows, it is possible to deploy the tools that embodied cognitive science has been developing these past decades to make nuanced analysis of sociopolitical urgent issues. Following Maiese and Hanna's (Maiese & Hanna, 2019) analysis of the nature of the cognitive institutions in which agents' lives unfold and their potential destructive and disempowering character, or the analysis of Slaby of the "corporate life's hack" (Slaby, 2016), our analysis of how digital platforms shape our habits allows us to consider these new cognitive ecologies (Smart et al., 2017) as a potential case of destructive institutions of contemporary cognitive capitalism. Furthermore, by the very nature of our predicament, current analyses should embrace also the collaboration with and the development of a political philosophy of technology and AI (Coeckelbergh, 2022; Hui, 2016; Lazar, 2023). A fruitful political analysis rooted in cognitive science and philosophy of technology allows us to identify the dangers of the attention economy upon the personal autonomy of its inhabitants.

One classical *topoi* of political philosophy is that of voluntary servitude: why human beings fight for their chains as if their salvation depended on it. In the control societies of cognitive capitalism, the chains take the form of screens, platforms and algorithms that capture our attention, organizing social and economic processes while affecting our well-being and autonomy. The digital panopticon, deployed today in "free" networks in the form of ambient computing, not only persists in the form of "choreographed" (Coeckelbergh, 2022) platform confessional machines, but has itself morphed into an economic digital engine fabricating *dividual*s regulated and governed according to protocols, algorithms and databases property of large technological companies. Our analysis allows formulating the phenomena in some of its hidden complexity. Our discussion of how habits become induced and stabilized and how the emergent networks of habits guide behaviour deeper into the same patterns, allows us to understand why sometimes we seem to be the initiators of our engagement with these digital chains, and why this doesn't strip them from their potential harmful character. Furthermore, it emphasizes the relevance of bringing into the scene the technopolitical milieu orchestrating this form of *dividual* governance.

Political critique must then pass through the opening of the black boxes that organize productive, social and subjective processes in the attention economy, on which our understanding of the civilisational crisis in which we find ourselves depends, ranging from the subsoil of the earth to the upper layers of the atmosphere, passing through the damaged subjectivities that populate the privileged developed societies. The cognitive sciences can help in this process by highlighting how our cognitive capacities are undermined by participating in the extractivist logics that organize contemporary societies. An approach such as the one we have proposed is capable of capturing the place that the digital platforms of the attention economy play in the processes of generation and destruction of subjectivity, identifying the perverse mechanisms that cognitive capitalism sets in motion by integrating our affective and communicative capacities into the processes of economic valorisation. Only by recovering collective autonomy in the design of techno-ecological environments can we recover sensorimotor and personal autonomy.

We don't want to end this paper on a gloomy note. Enactive theory, it must be underwritten, is not techno-deterministic. Following the importance of becoming in enactive theory, we have emphasized that humans and technologies must not be conceptualized as separate entities, but understood as interdependent and in a reciprocal constitutive relation; historical contingent assemblages emerging and evolving together. From this standpoint, it must be emphasized that technology has been omnipresent in the historical development of human capacities, and as such it hasn't been (can't be) neutral, shaping and being shaped by the values, beliefs and practices of its users and designers. In a society with evident asymmetries of power between agents and groups, this results in a cascade of effects arising from the resulting struggles that involve the production and implementation of new technologies.

This enactive insight opens the possibility of an emancipatory political perspective. Our analysis allows us to conclude that the design of our digital environments needs to be crucially participative (see the developments already made in the tradition of 'participatory design', Bødker et al., 2022), so that technologies are developed in line with the needs and values of the people that will use it. The impact of technology on society and the individuals is not predetermined but rather is the result of the dynamic interaction between the technology, its users and the socioeconomic and political context. The design of technology brings high political stakes. Technology should be built (and used) to enhance human autonomy and promote human flourishing, and to create a more just and equitable society (for a concrete example, see Barandiaran et al., 2024).

The scientific and theoretical knowledge we build is part of the education of attention with which we enter the digital world. Contemporary cognitive science can and should benefit from a conception of attention that doesn't forget the embodied and historical nature of our mental life. Cognitivist assumptions only get us, in the best scenario, to an abstract understanding of attention that hides the specific extractive, political and economical nature of our digital environment. An improved understanding of attention, grounded on the enactive notion of habit and its conception of agents as bundles of habits, allowed us to perceive the multiple dimensions concealed in the phenomena of attention and attention capture. Crucially, the normative dimension that the enactive framework provides makes it possible to connect these insights with the analysis of the harmful techno-political nature of algorithmic governance in contemporary platform capitalism. In this paper we have prompted (theoretical) attention to the consequences of different scientific theories of attention not only to correctly characterize contemporary techno-political economy and its psychological consequences, but also to move towards creating more sustainable attention *ecologies*, caring for the fragile and precarious nature of our identities as autonomous bundles of habits.

## Acknowledgements

MPV and XEB carried out this work in the context of the Outonomy Research Project, grant PID2019-104576GB-I00 funded by MCIN/AEI/10.13039/501100011033. Within this project, MPV is developing her PhD thanks to the grant PRE2020-096494 of the Spanish National Research Agency (AEI) and co-funded by the European Social Fund. Both authors have also benefited from being a member of the IAS-Research group, funded by the Basque Government (grant IT16668-22), PI Jon Umerez.## References

Aagaard, J. (2015). Drawn to distraction: A qualitative study of off-task use of educational technology. *Computers & Education, 87*, 90–97. http://dx.doi.org/10.1016/j.compedu.2015.03.010

Allport, A. (1987). Selection for action: Some behavioral and neurophysiological considerations of attention and action. In H. Heuer & A. Sanders, *Perspectives on perception and action* (Psychology Library Editions: Cognitive Science, pp. 395–420). Routledge.*Attention is all they need*    22